\documentclass[aps,prl,reprint,superscriptaddress,showpacs]{revtex4-1} 
\usepackage{graphicx}  
\usepackage{dcolumn}   
\usepackage{amssymb}
\usepackage{physics}
\usepackage{natbib}
\usepackage{color}
\usepackage[dvipsnames]{xcolor}
\usepackage[colorlinks=true, citecolor={blue!80!black}, urlcolor={blue!50!black}, linkcolor = {blue!80!black}]{hyperref}

\begin{document}
\title{Andreev Modes from Phase Winding in a Full-shell Nanowire-based Transmon}
\author{A.~Kringh{\o}j}
\affiliation{Center for Quantum Devices, Niels Bohr Institute,
University of Copenhagen, 2100 Copenhagen, Denmark}
\affiliation{Microsoft Quantum Lab-Copenhagen, Niels Bohr Institute,
University of Copenhagen, 2100 Copenhagen, Denmark}
\author{G.~W.~Winkler}
\affiliation{Microsoft Quantum, Station Q, University of California, Santa Barbara, California 93106-6105, USA}
\author{T.~W.~Larsen}
\author{D.~Sabonis}
\author{O.~Erlandsson}
\affiliation{Center for Quantum Devices, Niels Bohr Institute,
University of Copenhagen, 2100 Copenhagen, Denmark}
\affiliation{Microsoft Quantum Lab-Copenhagen, Niels Bohr Institute,
University of Copenhagen, 2100 Copenhagen, Denmark}
\author{P.~Krogstrup}
\affiliation{Center for Quantum Devices, Niels Bohr Institute,
University of Copenhagen, 2100 Copenhagen, Denmark}
\affiliation{Microsoft Quantum Materials Lab-Copenhagen, 2800 Lyngby, Denmark}
\author{B.~van~Heck}
\affiliation{Microsoft Quantum Lab Delft, Delft University of Technology, 2600 GA Delft, The Netherlands}
\author{K.~D.~Petersson}
\author{C.~M.~Marcus}
\affiliation{Center for Quantum Devices, Niels Bohr Institute,
University of Copenhagen, 2100 Copenhagen, Denmark}
\affiliation{Microsoft Quantum Lab-Copenhagen, Niels Bohr Institute,
University of Copenhagen, 2100 Copenhagen, Denmark}
\begin{abstract}
We investigate transmon qubits made from semiconductor nanowires with a fully surrounding superconducting shell. In the regime of reentrant superconductivity associated with the destructive Little-Parks effect, numerous coherent transitions are observed in the first reentrant lobe, where the shell carries 2$\pi$ winding of superconducting phase, and  are absent in the zeroth lobe. As junction density was increased by gate voltage, qubit coherence was suppressed then lost in the first lobe. These observations and numerical simulations highlight the role of winding-induced Andreev states in the junction.
\end{abstract}

\maketitle

In Josephson junctions formed by two superconductors separated by a coherent transmitting region, multiple electron-hole reflections at the superconductor boundaries result in a discrete subgap spectrum of Andreev bound states (ABSs) whose energy depends on the difference in phase, $\delta\varphi$, across the junction~\cite{andreev_1964}. Recent microwave measurements have explored subgap Andreev spectra ~\cite{bretheau_2013, woerkom_2017} and coherence~\cite{janvier_2016, Hays_2018}, including effects of spin splitting and spin-orbit coupling in nanowire (NW) junctions~\cite{Tosi_2019, hays_2020}. 

Extending this development, hybrid semiconductor-superconductor NWs have been used to realize transmonlike qubits~\cite{larsen_2015, delange_2015}, operated at large ratios of the Josephson energy $E_J$ to the charging energy $E_C$~\cite{koch_2007}, where $\delta\varphi \sim 0$.
The gate-tunable NW junctions used in these devices typically have a few channels with high transparency~\cite{Spanton_2017, Goffman_2017, Anharmonicity}, which has observable consequences on qubit properties such as anharmonicity~\cite{Anharmonicity} and charge dispersion~\cite{arno,dispersion}.
However, ABSs themselves are not readily observed in transmon measurements because near $\delta\varphi = 0$ transition frequencies are often much higher than the qubit frequency and well outside of the usual operational bandwidth (2--10~GHz).
Tuning the phase difference near $\delta\varphi = \pi$ is usually required to lower ABS transitions to a measurable range~\cite{janvier_2016, Hays_2018}.

\begin{figure}[b]
    \centering
        \hspace{-2mm}\includegraphics[width=1\columnwidth]{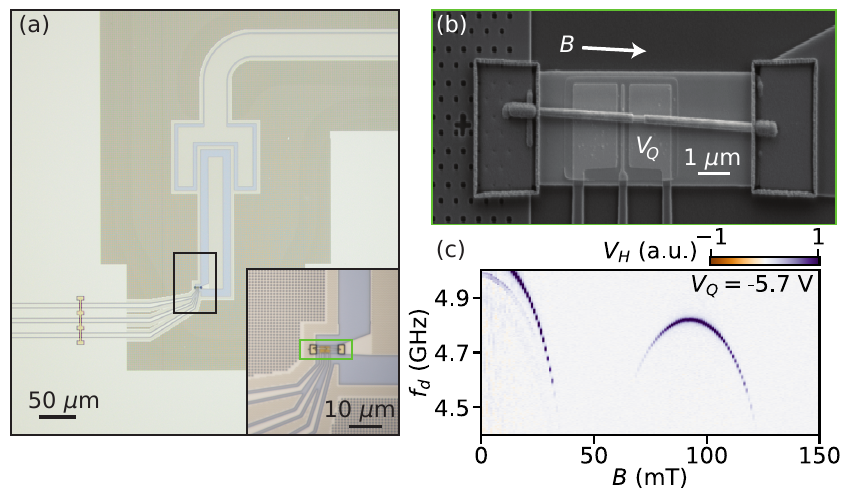}\vspace{-4mm}
    \caption[Field compatible device and Little-Parks effect]{(a) Optical micrograph of the qubit on $\sim1$~mm and $\sim0.1$~mm scale (inset) for device 1. The readout resonator is capacitively coupled to the rectangular qubit island, with an InAs/Al nanowire (NW) at the bottom (black rectangle). Flux-pinning holes are patterned near the island and resonator. (b) Electron micrograph of the NW [green rectangle in (a) inset], with junction controlled by gate voltage $V_Q$. Other electrodes not used. A magnetic field $B$ is applied parallel to the NW. (c) Demodulated transmission voltage $V_H$ as a function of $B$ and qubit drive frequency $f_d$, showing reentrant qubit frequency $f_{01}$ (device~1). Line median subtracted from each column. 
    }
    \label{fig:Device and sketch}\vspace{-4mm}
\end{figure}

Here, we investigate low-energy ABSs in superconducting full-shell NW-based gatemons, which exhibit a destructive Little-Parks effect~\cite{LittleParks, sabonis2020}. An applied magnetic field parallel to the NW induces a sequence of reentrant superconducting lobes, each associated with a different winding number, $n$, of the superconducting phase around the shell~\cite{Sole_Georg_2020, sole_2020}. ABSs appear in the low-energy spectrum of the device in the first lobe ($n=1$) and are absent in the zeroth lobe ($n=0$), consistent with numerical simulations, also presented here.

Nanowire-based qubits in a circuit quantum electrodynamics~\cite{wallraff_2004, blais_2004} architecture were fabricated on high resistivity silicon substrates covered with a 20~nm thin NbTiN film~\cite{lead}. Superconducting qubit islands, $\lambda/4$ distributed readout resonators with resonance frequencies $f_\text{res}\sim5$~GHz, transmission line, junction gates, and on-chip gate filters were defined with electron-beam lithography followed by reactive ion etching; see Fig.~\ref{fig:Device and sketch}(a).
NWs were placed on bottom-gates separated by a 15~nm thin HfO$_2$ dielectric.
The NWs consisted of an InAs core of $140$~nm in diameter, fully coated with $35$~nm epitaxial Al~\cite{krogstrup_2015}.
By a selective wet etch, a $\sim400$~nm segment of the Al shell was removed, creating a Josephson junction (JJ) [Fig.~\ref{fig:Device and sketch}(b)]. Connecting superconducting leads to ground and qubit island completed the gatemon circuit~\cite{larsen_2015, delange_2015}.
The ground plane was patterned with flux-pinning holes, crucial for magnetic field compatibility of the readout resonators~\cite{kroll_2018, kroll_2019}.
Measurements are presented for two devices, denoted 1 and 2.

The qubit frequency $f_{01}$ is measured as a function of magnetic field, $B$, applied parallel to the NW using both two-tone spectroscopy and single-tone spectroscopy. In two-tone spectroscopy, a pulsed qubit drive tone of variable frequency $f_d$ is followed by a pulsed tone at a fixed readout frequency. By measuring the demodulated heterodyne transmission voltage $V_H$, the qubit frequency, $f_{01}$, can be inferred from the dispersive interaction between the qubit and resonator~\cite{wallraff_2004, blais_2004}. We measure both the in-phase ($I$) and quadrature ($Q$) components of the transmission signal, defining $V_H$ by rotating the data in the $IQ$-plane to maximize the signal along the real axis~\cite{supplement, code}. Two-tone measurement of the qubit frequency spectrum as a function of $B$~\cite{sabonis2020}, through the zeroth and first lobes ($n =$ 0, 1) is shown in Fig.~\ref{fig:Device and sketch}(c).

\begin{figure}
    \centering
        \hspace{-2mm}\includegraphics[width=1\columnwidth]{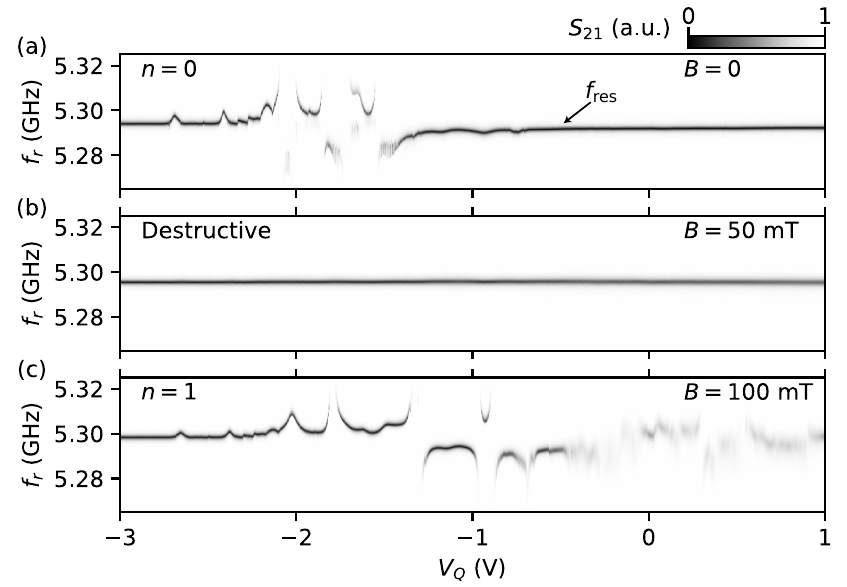}\vspace{-4mm}
    \caption[Damping of the resonator]{Single-tone spectroscopy using transmission voltage amplitude $S_{21}$ as a function of gate voltage $V_Q$ and resonator drive frequency $f_r$ in the zeroth lobe, destructive regime, and first lobe, for device 2. (a) In the zeroth lobe ($B=0$, $n=0$), the resonance frequency $f_\text{res}$ (arrow) is modulated by $V_Q$ due to the varying qubit frequency. (b) In the destructive regime ($B=50$~mT), the resonator shows no dependence on $V_Q$. (c) In the first lobe ($B=100$~mT, $n=1$), the resonator shows gate dependence similar to (a) for $V_Q\lesssim-1$~V, and a broadened line, indicating increased damping, for $V_Q\gtrsim-1$~V.
    }
    \label{fig:Damping}\vspace{-4mm}
\end{figure}

Single-tone spectroscopy directly measures the modulation of the resonance frequency $f_\text{res}$ of the resonator due to its interaction with the qubit. Figure~\ref{fig:Damping} shows the transmission voltage amplitude $S_{21}$ as a function of the resonator drive frequency $f_r$. In the zeroth lobe ($B=0$), we observe a nonmonotonic modulation of $f_\text{res}$ associated with the voltage modulation of $f_{01}$ as $V_Q$ is increased from complete depletion at $V_Q\approx-3$~V [Fig.~\ref{fig:Damping}(a), $n=0$].
For $V_Q\gtrsim-2$~V, several avoided crossings are observed, indicating that the qubit is tuned in and out of resonance with the resonator, as frequently observed for gatemon qubits~\cite{larsen_2015, delange_2015}.
For $V_Q\gtrsim-1$~V, $f_\text{res}$ approaches its unshifted value, indicating a vanishing dispersive shift due to $f_{01}$ being far above $f_\text{res}$. In the destructive regime [$B=50$~mT, Fig.~\ref{fig:Damping}(b)], $f_\text{res}$ shows no dependence on $V_Q$, as expected at flux $\Phi\sim\Phi_0/2$, where superconductivity in the Al shell is lost.
In the first lobe [$B=100$~mT, Fig.~\ref{fig:Damping}(c)], for $V_Q\lesssim-1$~V, $f_\text{res}$ yields similar modulation compared to $n=0$. This suggests a similar $V_Q$ dependence of the qubit for $n=1$.
For $V_Q\gtrsim-1$~V, the spectrum is strikingly different from that measured in the zeroth lobe: the position of the resonant dip is subject to more fluctuations, and its width is increased.

We interpret the damping of the resonator as an increased decay rate caused by the qubit, indicating that the relaxation rate of the qubit is much larger than that of the resonator.
We emphasize that the onset of enhanced relaxation in Fig.~\ref{fig:Damping}(c) is not gradual in $B$ but occurs suddenly in the first lobe.
This is demonstrated by a series of identical measurements taken at $5$~mT intervals in both lobes~\cite{supplement}, suggesting that the loss of qubit coherence for increasing $V_Q$ is a phenomenon associated with $n=1$, but not $n=0$.

\begin{figure}
    \centering
        \hspace{-2mm}\includegraphics[width=1\columnwidth]{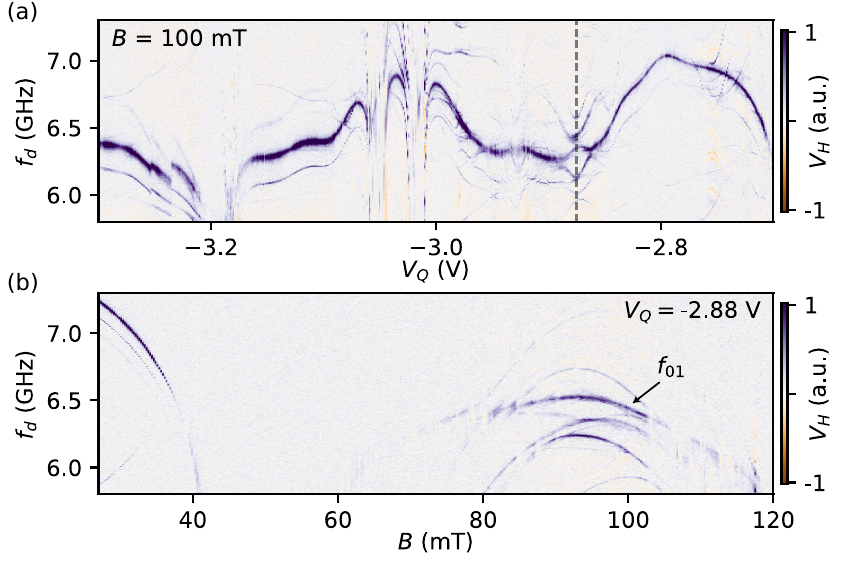}\vspace{-4mm}
    \caption[Measurements of phase-twisted Andreev transitions]{(a) Two-tone spectroscopy shows demodulated transmission $V_H$ as a function of gate voltage $V_Q$ and qubit drive frequency $f_d$ in the first lobe ($B=100$~mT) for device 1. Numerous additional transitions are seen near the main qubit transitions ($f_{01}$ and higher harmonics). (b) $V_H$ as a function of $B$ and $f_d$ at $V_Q=-2.88$~V [dashed line in (a)]. Qubit transition and additional transitions show reentrance from destructive Little-Parks effect. In the zeroth lobe only the main qubit transitions are seen. The qubit frequency $f_{01}$ (arrow) is a associated
with the transition that persist as a function of $V_Q$ in (a). We note a mirrored qubit transition peak for $V_Q\lesssim-3.1$~V and $V_Q\gtrsim-2.8$~V in (a) and for $B<40$~mT in (b), associated with sideband leakage of the pulse modulation. Line median subtracted from each column.
    }
    \label{fig:Fireworks}\vspace{-4mm}
\end{figure}

In the regime just before the onset of increased resonator relaxation, we directly map the qubit dependence of $V_Q$ in the first lobe ($B=100$~mT), by two-tone spectroscopy; see Fig.~\ref{fig:Fireworks}(a).
Here, a range of unconventional energy transitions emerge. These transitions show strong gate dependence near certain values of $V_Q$, and several avoided crossings with the qubit transition, indicating that the transitions couple to the qubit.
These transitions are observed for $n=1$, clearly deviating from the spectra for $n=0$~\cite{supplement}. For $V_Q<-3.3$~V, only the usual gatemon transition frequencies ($f_{01}$ and higher harmonics) are observed~\cite{note_opposite, supplement}. For $V_Q\gtrsim-2.7$~V qubit coherence is lost and the qubit frequency is not observed~\cite{supplement}. We associate the emergence of transitions and the loss of coherence with the increasing density of low-energy sub-gap states, which leads to a heavily damped qubit transition that in turn cause the damping of the readout resonator via the qubit-resonator coupling. The measurements in Fig.~\ref{fig:Damping} and Fig.~\ref{fig:Fireworks} are from two different devices with different threshold $V_Q$ at which coherence is lost. The characteristic loss of coherence is the same in both devices.

The unconventional transition frequencies are resolvable throughout the first lobe ($B\sim70$--$120$~mT), while not being visible in the zeroth lobe for $B\lesssim40$~mT, as shown in Fig.~\ref{fig:Fireworks}(b). This demonstrates that the emergence of these transitions is a distinct feature associated with $n=1$, but not $n=0$.
These transitions both exhibit local maxima at larger values of $B$ than $f_{01}$, and a distinctively different dependence on $B$, compared to $f_{01}$ [arrow in Fig.~\ref{fig:Fireworks}(b)]. 

We perform simulations of a full-shell NW junction of similar dimensions as the measured devices.
A hexagonal InAs/Al NW is modelled with the same methods and parameters as in Ref.~\cite{Sole_Georg_2020}, here adapted to the JJ geometry. 

\begin{figure}
    \centering
        \hspace{-2mm}\includegraphics[width=1\columnwidth]{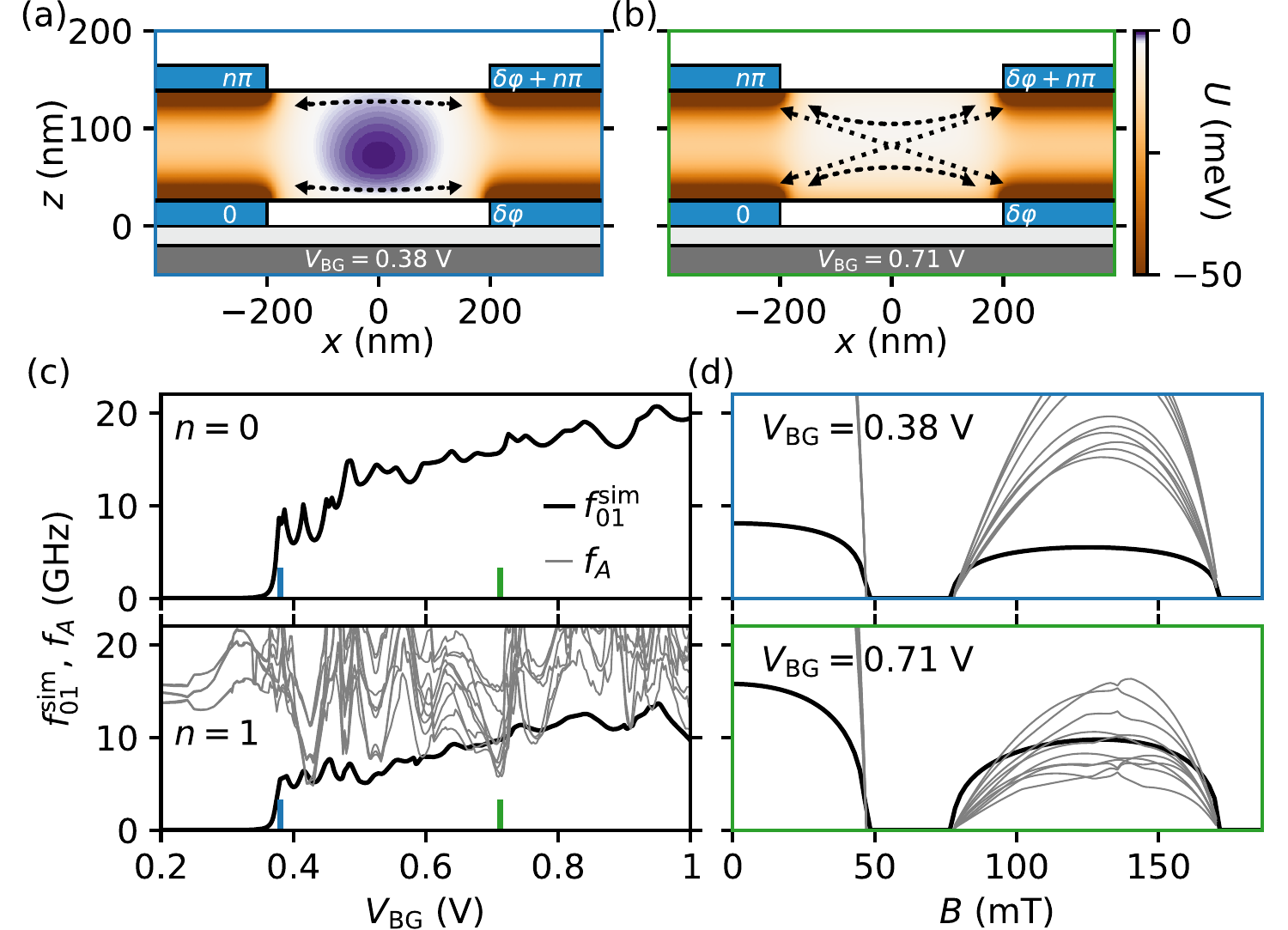}\vspace{-4mm}
    \caption[Numerical simulations of qubits excitations]{Simulated electrostatic potential $U$ near the InAs NW junction, between full-shell Al regions (blue) (a) for a nearly depleted junction, for voltage $V_\text{BG}=0.38\,$V on the back-gate (gray) separated by HfO$_2$ dielectric (light gray) and (b) for a more populated junction, $V_\text{BG}=0.71\,$V. For $V_\text{BG}=0.38\,$V, the center region of the junction is depleted (purple). Superconducting phases are indicated in the shell regions. Dotted arrows illustrate relevant trajectories across the junction. (c) Simulated qubit frequency $f_{01}^{\rm sim}$ (black) and two-quasiparticle transition frequencies $f_A$ at $\delta\varphi=0$ (gray) as a function of $V_\mathrm{BG}$ in the zeroth lobe ($B=0$, $n=0$, top panel) and in the first lobe (center of the lobe, $n=1$, bottom panel). Two-quasiparticle transitions are off the scale shown for $n=0$. (d) $f_{01}^{\rm sim}$ and $f_A$ as a function of $B$ for $V_\text{BG}=0.38\,$V (top panel) and $V_\text{BG}=0.71\,$V  (bottom panel).  
    }
    \label{fig:theory lobe}\vspace{-4mm}
\end{figure}

As shown in Ref.~\cite{Sole_Georg_2020}, the full-shell NW can host a topological or trivial phase for $n=1$. Since the two phases
cannot be distinguished purely from the excited ABS energies around $\delta\varphi=0$, the presented results are obtained with parameters corresponding to the trivial phase, although the simulated ABS spectrum in the topological regime is qualitatively similar at $\delta\varphi\sim0$~\cite{supplement, footnote_topo}. All simulations assume a band offset between InAs and Al of 140~meV and no spin-orbit coupling $\alpha=0$. 
Figures~\ref{fig:theory lobe}(a,~b) show longitudinal cuts of the simulated JJ with the electrostatic potential inside the InAs NW, which illustrate the formation of a depletion region for low back-gate voltages, occurring around $V_\mathrm{BG}=0.38$~V, unlike an open junction at a representative $V_\mathrm{BG}=0.71$~V.

To obtain a qualitative comparison with the experiment, we perform numerical simulations of the supercurrent using the Kwant package~\cite{kwant} and the analysis developed in Refs.~\cite{Ostroukh_2016,Zou_2017}.
From the critical current $I_c$, the simulated qubit frequency $f_{01}^{\rm sim}=\sqrt{8 E_J E_C}/h=\sqrt{2E_CI_c/h\pi e}$ is obtained, where $E_C/h\sim 500$~MHz is estimated from electrostatic simulations of the qubit island~\cite{comsol} and $E_J=\hbar I_c/2e$. 
Figure~\ref{fig:theory lobe}(c) shows $f_{01}^{\rm sim}$ for $n=0$ and $n=1$ as a function of $V_\mathrm{BG}$.
While $f_{01}^{\rm sim}$ in the first lobe is reduced compared to the zeroth lobe, the general shape are similar between the lobes. This is also observed experimentally; see Fig.~\ref{fig:Damping}. We interpret this reduction of $f_{01}^{\rm sim}$ as resulting from the overall reduced gap for $n=1$, as discussed previously in Ref.~\cite{sabonis2020}.

By calculating the ABS transition energies from the density of states~\cite{supplement}, the excitation frequencies $f_A$ of energy states with two excited quasiparticles at $\delta\varphi=0$ are identified and shown in Figs.~\ref{fig:theory lobe}(c)~and~\ref{fig:theory lobe}(d) along with $f_{01}^{\rm sim}$.
When these transitions are near the qubit transition, they are expected to be activated in spectroscopic measurements via the coupling to the plasma mode~\cite{keselman_2019} (coupling not included in our simulations).
In the zeroth lobe all energies of two-quasiparticle states are well above the qubit frequency.
However, in the first lobe these states are much lower in energy and $f_A$ even cross the qubit frequency for several transitions for certain gate voltages as the junction is opened for increasing $V_\mathrm{BG}$. This is in qualitative agreement with the experimental observations shown in Fig.~\ref{fig:Fireworks}(a). 

Figure~\ref{fig:theory lobe}(d) shows the magnetic field dependence of $f_{01}^{\rm sim}$ and $f_A$ for two different $V_\mathrm{BG}$, representing a junction near depletion (top panel) and in an open regime (bottom panel). 
$f_A$ for all two-quasiparticle transitions show a roughly parabolic dependence, distinctively different from $f_{01}^{\rm sim}$. This is consistent with the expected scaling of two-quasiparticle state energies with $\Delta$~\cite{beenakker_1991}, whereas $f_{01}^{\rm sim}$ is expected to scale with $\sqrt{\Delta}$ (for $f_{01}^{\rm sim}=\sqrt{8 E_J E_C}/h$~\cite{koch_2007} and $E_J\propto\Delta$). These expectations are also consistent with the different field dependencies of the transitions, observed experimentally in Fig.~\ref{fig:Fireworks}(b).
Furthermore, the local maxima of the simulated two-quasiparticle transitions are typically shifted to higher magnetic fields than the local maximum of $f_{01}^{\rm sim}$, which is similar to the experimental observations shown in Fig.~\ref{fig:Fireworks}(b). This suggests that the wave functions of the associated states have a smaller effective cross-section smaller than the Al shell and thus are pierced by a correspondingly smaller flux. 

\begin{figure}
    \centering
        \hspace{-2mm}\includegraphics[width=1\columnwidth]{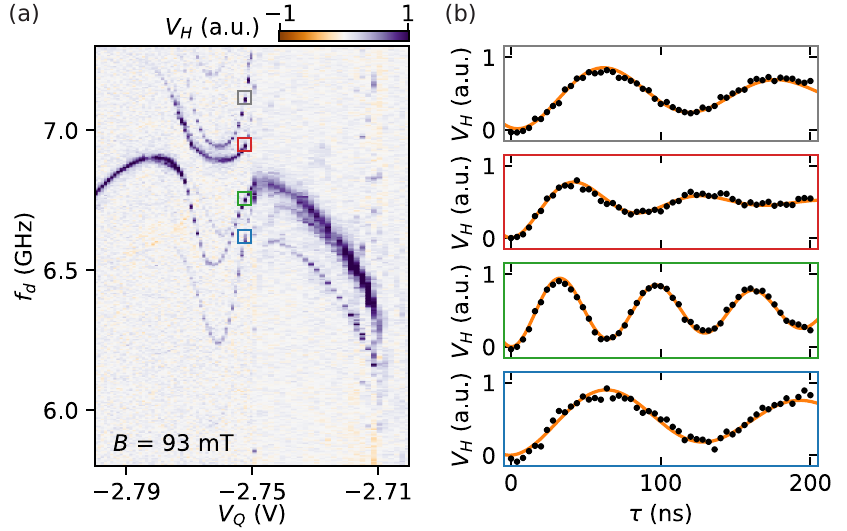}\vspace{-4mm}
    \caption[Time domain measurements of Andreev transitions]{(a) Demodulated transmission $V_H$ as a function of $B$ and qubit drive frequency $f_d$ in a narrow range of $V_Q$ in the first lobe ($B=93$~mT) for device 1. The Andreev transitions are strongly gate dependent with local minima at similar $V_Q$. The drive power was increased by 10~dBm at $V_Q>-2.75$~V to ensure visibility of the qubit transition until coherence was lost for $V_Q\gtrsim-2.7$~V. Line median subtracted from each column. (b) Rabi measurements of $V_H$ as a function of varying drive pulse width $\tau$ at $V_Q=-2.752$~V [colored squares in (a) match frame colors in (b)]. All transitions yield coherent Rabi oscillations, with the fastest oscillations of the transitions nearest the uncoupled qubit transitions (green and red frames). The experimental data (black data points) are fitted to exponentially damped sinusoids (orange curves) and normalized to the extracted fit parameters.
    }
    \label{fig:gate and time domain}\vspace{-4mm}
\end{figure}

The qualitative agreement between experiments and numerics suggests that the visible transition frequencies are explained by two-quasiparticle excitations of ABSs in the junction that emerge due to the non-trivial phase winding in the first lobe. When $n=1$, the flux-induced winding causes a circumferential dependence of the phase of the order parameter on both sides of the junction.
Therefore, ABSs can experience multiple different phase differences depending on the particular trajectory they travel across the junction.
In particular, for $n=1$, trajectories traveling diagonally across the junction would experience a $\pi$-phase shift at $\delta\varphi=0$ and thus cause the presence of low-energy ABSs.
These trajectories are more likely to occur as $V_\mathrm{BG}$ is increased and the potential barrier is reduced [Figs.~\ref{fig:theory lobe}(a) and \ref{fig:theory lobe}(b)], which lowers the ABS energy at $\delta\varphi=0$, resulting in an increasing density of low-energy states (see also~\cite{supplement}). We attribute the transitions near the qubit frequency, observed both experimentally (Fig.~\ref{fig:Fireworks}) and numerically (Fig.~\ref{fig:theory lobe}), to result from this increasing density of low-energy sub-gap ABSs for $n=1$.
Additionally, the emergence of these transitions immediately before the experimental observation of the increased relaxation of the resonator [Fig.~\ref{fig:Damping}(c)] suggests that the increased relaxation also occurs as a result of the increasing low-energy sub-gap density of states.

We focus on a narrow representative region of $V_Q$ [Fig.~\ref{fig:gate and time domain}(a)] to probe the coherence properties of the transitions in the first lobe. 
Throughout this regime the transitions can be driven coherently with examples of Rabi measurements across the transitions at $V_Q=-2.752$~V shown in Fig.~\ref{fig:gate and time domain}(b).
All transitions yield $T_1$-relaxation times of 3--5~$\mu$s, measured in a subsequent measurement.
It is observed that the transitions closest to the uncoupled qubit transition yield the fastest oscillations at constant drive power [green and red panels in Fig.~\ref{fig:gate and time domain}(b)]. Transitions originating from Andreev states are expected to be visible via the coupling to the qubit~\cite{keselman_2019}, where the microwave coupling is expected to depend on the detuning of the Andreev transitions to $f_{01}$. This is consistent with the variations in the Rabi frequencies shown in Fig.~\ref{fig:gate and time domain}(b). This supports the conclusion that the observed coherent energy transitions originate from Andreev bounds states due to the non-trivial phase winding in the first lobe. 

These results motivate future experiments investigating new regimes for Andreev qubits, utilizing the direct inductive coupling to the bound state spectrum~\cite{janvier_2016, Hays_2018}. Directly probing the Andreev spectrum may allow experiments to investigate the predicted emergence of topological regimes~\cite{Sole_Georg_2020}, in particular for shorter junctions with a lower density of Andreev bound states. 

\begin{acknowledgments}
We thank Andrey Antipov, Bela Bauer, Lucas Casparis, Anna Keselman, and Ivana Petkovic for valuable discussions. 
We acknowledge Marina Hesselberg, Karthik Jambunathan, Robert McNeil, Karolis Parfeniukas, Agnieszka Telecka, Shivendra Upadhyay, and Sachin Yadav for the device fabrication.  Research was supported by Microsoft, the Danish National Research Foundation, and the European Research Council under grant HEMs-DAM No.716655.
\end{acknowledgments}

%

\newpage

\newcommand{\beginsupplement}{%
     
        \setcounter{figure}{0}
        \renewcommand{\thefigure}{S\arabic{figure}}%
     }
\newpage
\onecolumngrid
\section{Supplementary Material}
\maketitle
\beginsupplement
\subsection{Experimental setup}

\begin{figure}[!b]
    \centering
        \includegraphics[width=1\textwidth]{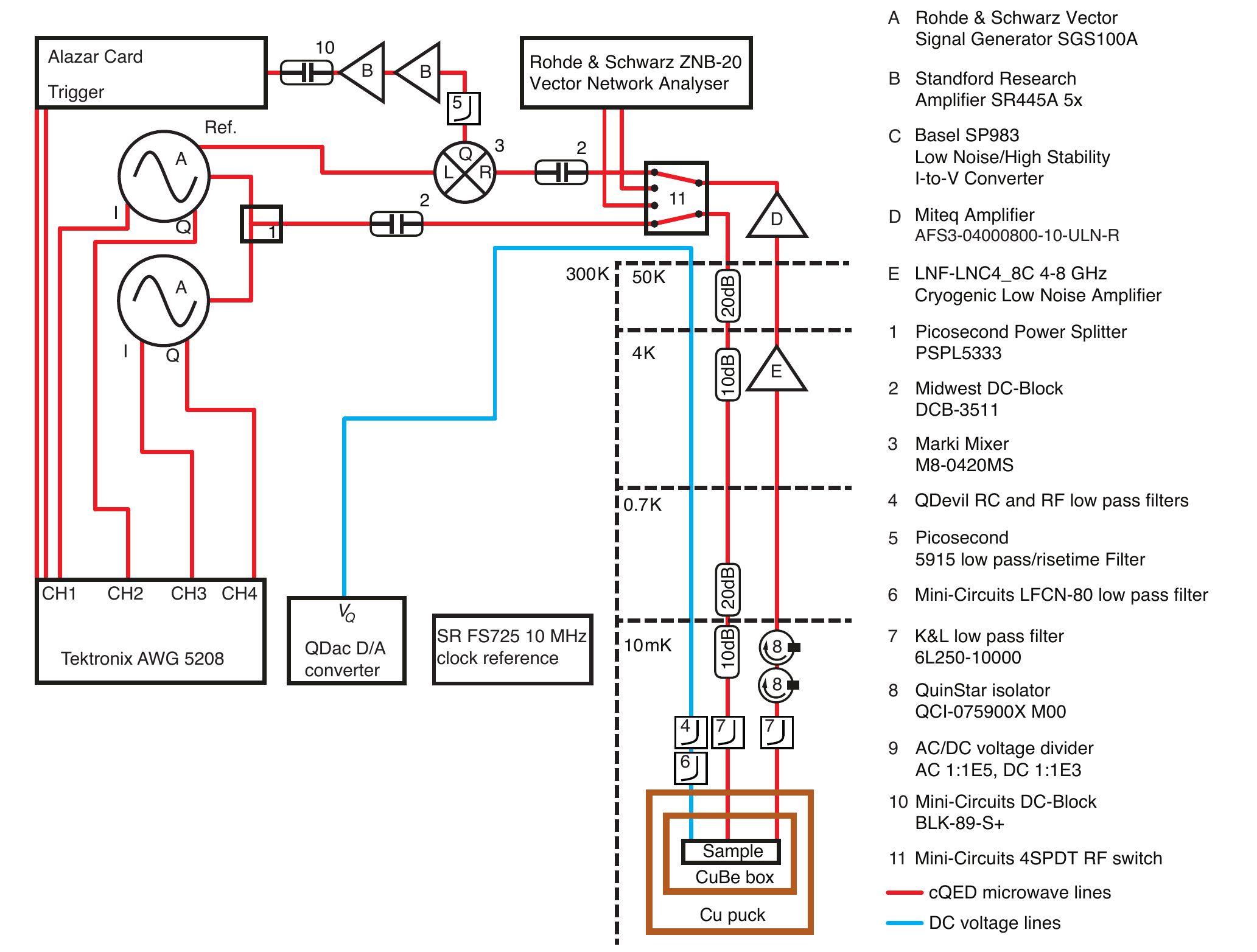}\vspace{-2mm}
    \caption{\vspace{-0mm} Detailed schematic of the experimental setup. Input microwave signal (red) is applied either by two combined RF signals or the VNA, controlled by an RF switch. The return signal is amplified at 4~K and at room temperature before going to the mixer and further demodulation or the VNA. DC lines (blue) connect the sample and are used to control the junction gate voltages $V_Q$. 
    }
    \label{fig:setup}
\end{figure}

Measurements were acquired in a cryofree dilution refrigerator with a base temperature of $\sim20$~mK. Figure~\ref{fig:setup} presents a detailed schematics of the experimental setup. The sample is mounted at the mixing chamber plate in a Cu circuit board mounted in a indium sealed CuBe box shielded with a second Cu box. Microwave signals (red) are routed to the sample via two microwave coaxial drive lines either from the vector network analyzer (VNA) or the demodulation circuit. The demodulation circuit consists of a mixer to down convert the output signal to an intermediate frequency by mixing with a reference tone followed by a low pass filter. This signal is then digitized and digitally down converted to record the magnitude and phase components of the readout signal. A second microwave source generates the qubit drive tones, applied via the same drive line. A Mini-Circuits RF switch allows to controllably switching between VNA measurements used to probe the resonator and the demodulation circuit used for pulsed AWG-based measurements.
The SR FS725 10~MHz clock reference synchronizes the Alazar card,
signal generators and the AWG.

\subsection{Field dependence of resonator frequency}

\begin{figure}
    \centering
        \includegraphics[width=1\textwidth]{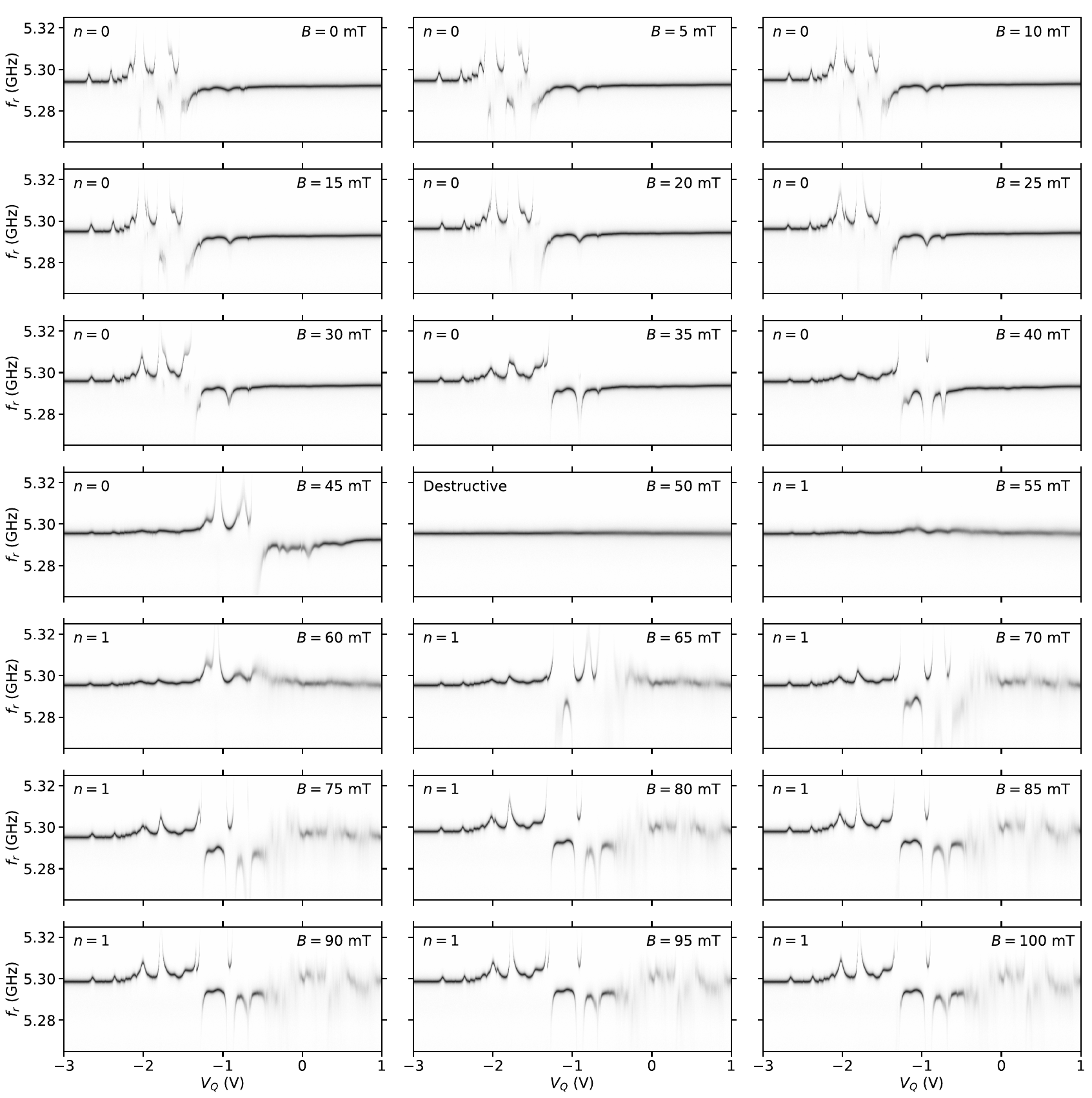}\vspace{-2mm}
    \caption[Full measurement of the lobe dependent resonator damping]{\vspace{-0mm} Single-tone spectroscopy using transmission voltage amplitude $S_{21}$ as a function of junction gate voltage $V_Q$ and drive frequency $f_d$ at increasing parallel magnetic field $B=0$--$100$~mT in steps of 5~mT. A characteristic damping of the resonator is observed for $V_Q\gtrsim-1$~V as the transition from the zeroth lobe ($n=0$) to the first lobe ($n=1$) occurs via the destructive regime.
    }
    \label{fig:full damping}
\end{figure}

To investigate the observed damping of the readout resonator, we systematically map $f_\text{res}$ for increasing values of $B$. We repeat the single-tone spectroscopy scans performed in Fig.~\ref{fig:Damping} in the main text for values of $B$ increasing by 5~mT, as shown in Fig.~\ref{fig:full damping}. It is observed that the overall spectrum remain roughly unchanged up to $B=45$~mT, except small changes in the avoided crossings, which is attributed to the field-induced decrease in $\Delta$. At $B=50$--$55$~mT, we enter the destructive regime ($\Phi\sim\Phi_0/2$), where superconductivity is destroyed. When entering the first lobe, superconductivity is restored. In the first lobe, the characteristic behavior of the spectra is very similar with few variations due to the field modulation of $\Delta$. We observe the damping of the resonator for all values of $B$ for $V_Q\gtrsim-0.5$~V, in contrast to the zeroth lobe.

\begin{figure}[!h]
    \centering
        \includegraphics[width=0.6\textwidth]{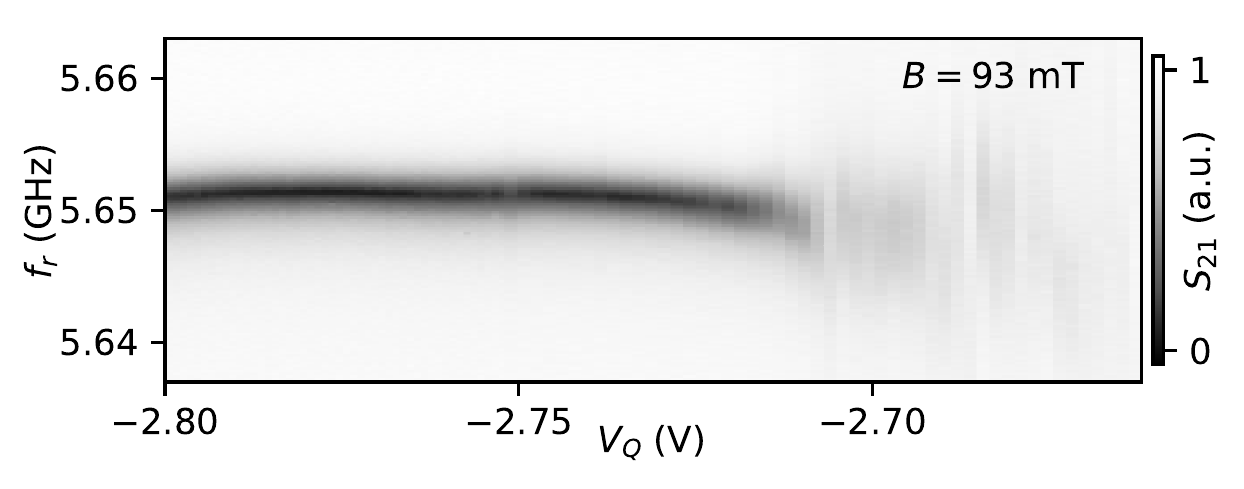}\vspace{-2mm}
    \caption[Corresponding resonator illustrating the damping effects]{\vspace{-0mm} Corresponding resonator scan as a function of $V_Q$ measured interleaved with the two-tone spectroscopy measurement presented in Fig.~\ref{fig:gate and time domain}(a) in the main text.
    }
    \label{fig:corresponding resonator damping}
\end{figure}

In Figs.~\ref{fig:Fireworks}, and~\ref{fig:gate and time domain} in the main text the behavior of the low-energy Andreev transitions are investigated. For $V_Q>-2.7$ qubit coherence is lost. Figure~\ref{fig:corresponding resonator damping} shows a resonator scan in this region, where we similarly observe the damping of the resonance frequency. This is consistent with the interpretation that the Andreev states are only visible in an open junction regime, with the resonator damping occurring due to softening of the superconducting gap. 

\subsection{Gate dependence in zeroth and first lobe}
\label{sec:gate comparison 0 and 1}
\begin{figure}[!h]
    \centering
        \includegraphics[width=0.6\textwidth]{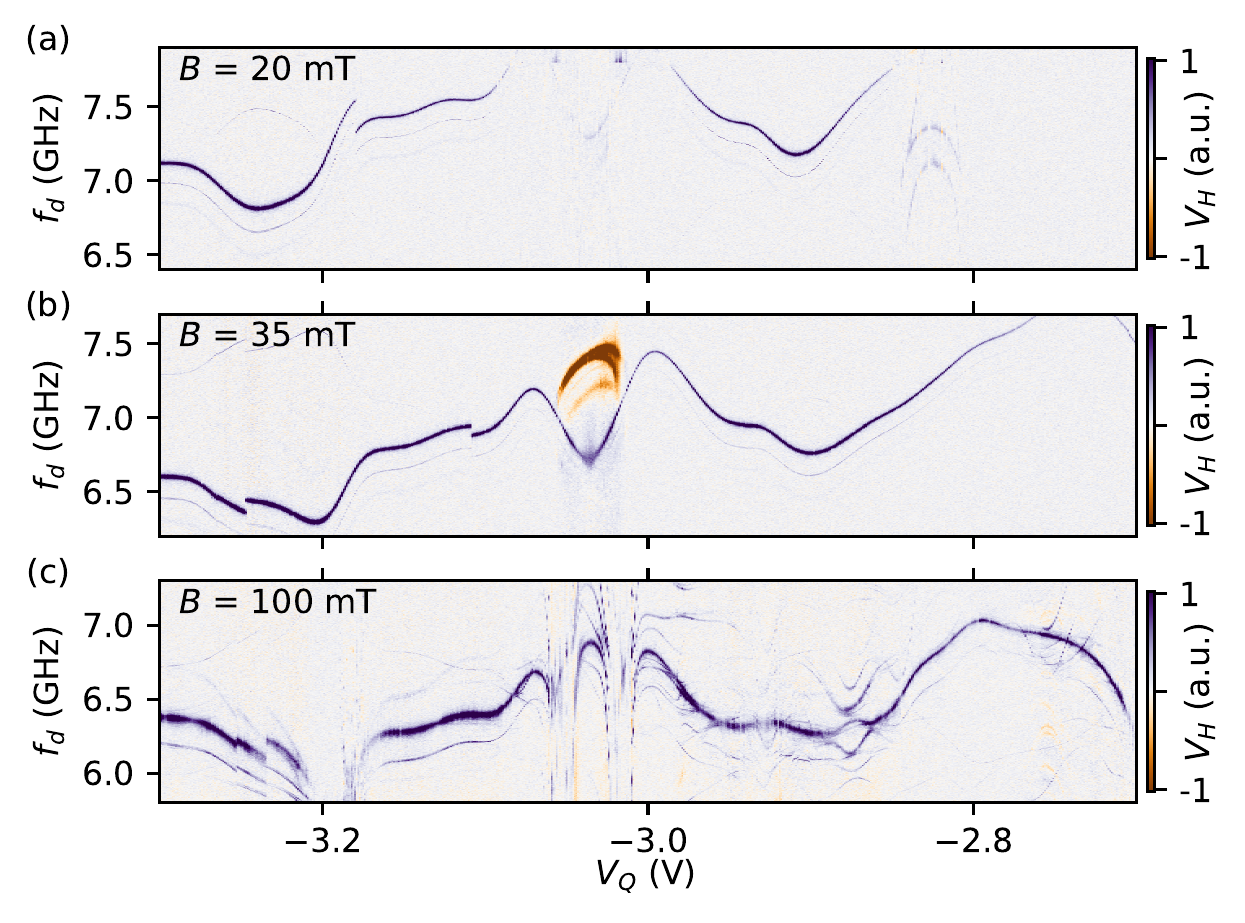}\vspace{-2mm}
    \caption[Comparison of 0th and 1st lobe two-tone spectroscopy]{\vspace{-0mm} Two-tone spectroscopy measurement as a function of qubit drive $f_d$ and $V_Q$ at $B=20$, $B=35$ and $100$~mT in (a), (b) and (c) to compare the gate dependence in the zeroth and first lobe. Panel (c) is also presented in Fig.~\ref{fig:Fireworks}(a) in the main text. Mirrored qubit transition for $V_Q\lesssim-3.1$~V due to sideband leakage. Line median subtracted from each column.
    }
    \label{fig:gate comparison 0th and 1st}
\end{figure}

To further support that the observed transitions are a characteristic phenomenon associated with phase twists in the first lobe, we repeat the gate scan shown in Fig.~\ref{fig:Fireworks}(a) in the main text at $B=20$ and $B=35$~mT, shown in Figs.~\ref{fig:gate comparison 0th and 1st}(a,~b). We observe a traditional gatemon spectrum, where the power broadened $0\to1$ and the two-photon $0\to2$ transitions are visible without all the additional transitions visible at $B=100$~mT [Fig.~\ref{fig:gate comparison 0th and 1st}(b)]. The absence of extra transitions lines in the zeroth lobe supports the interpretation of low-energy Andreev energy states due to the lobe dependent phase twists. 
We note the small region in Fig.~\ref{fig:gate comparison 0th and 1st}(a) near $V_Q\sim-2.8$~V and in Fig.~\ref{fig:gate comparison 0th and 1st}(b) near $V_Q\sim-3$~V, where extra transitions are visible. We speculate that these transitions arise from excited Andreev states, making them visible at low frequencies for $n=0$ also. 

\begin{figure}[!h]
    \centering
        \includegraphics[width=0.6\textwidth]{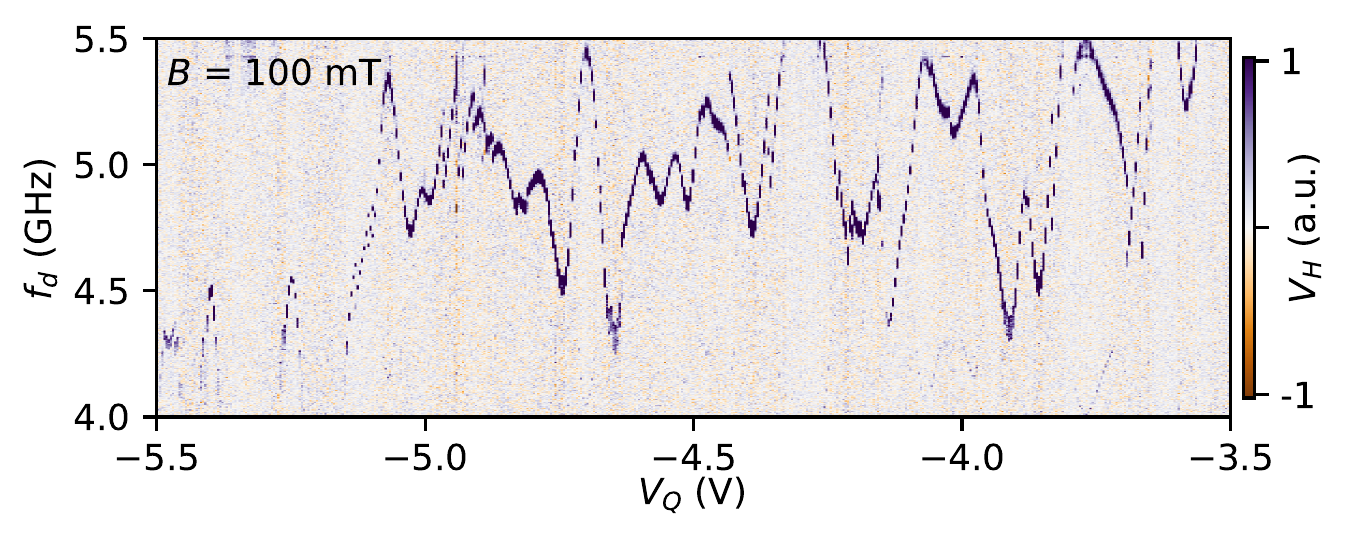}\vspace{-2mm}
    \caption[Two-tone spectroscopy measurements at low $V_Q$]{\vspace{-0mm} Two-tone spectroscopy measurement as a function of $f_d$ at $B=100$~mT for decreasing $V_Q$. In this regime, no additional transitions are observed. Mirrored qubit transition due to sideband leakage. Line median subtracted from each column.
    }
    \label{fig:gate low freq 100 mT}
\end{figure}

From the numerical modeling presented in Fig.~\ref{fig:theory lobe} in the main text, the observation of low-energy Andreev states are expected to be achievable for increasing values of $V_Q$. This interpretation is supported by gate scans probing the qubit spectrum for decreasing $V_Q$, as shown in Fig.~\ref{fig:gate low freq 100 mT}. Here, we map the qubit spectrum for values of $V_Q$ below the values shown in Fig.~\ref{fig:gate comparison 0th and 1st}, and no additional states are observed with only a single qubit transition line visible. This observation is consistent with creating a junction barrier for low values of $V_Q$, and the effect of the phase dependent junction paths are much less important. As a result the gatemon spectrum resembles the usual spectrum obtained in the zeroth lobe.
 
\subsection{Details of the numerical simulations}
\label{sec:num}

For details of the implementation of the numerical simulations see Ref.~\cite{Sole_Georg_2020}. The differences with respect to Ref.~\cite{Sole_Georg_2020} are the Josephson junction device geometry and that the simulations in the main text are performed in the topologically trivial regime of the full shell wire (band offset $U_0=140$\,meV and $\alpha=0$).

The local density of states (LDOS) is obtained from the Green's function,
\begin{equation}
G^{-1}(\omega)=(\omega+i\eta)\mathbb{I}-H_\mathrm{SM}-\Sigma^{(\mathrm{SC})}(\omega, \eta),
\end{equation}
where $\omega$ is the energy, $\eta$ the level broadening, $H_\mathrm{SM}$ the semiconductor Hamiltonian and $\Sigma^{(\mathrm{SC})}$ the self-energy of the integrated out superconductor Hamiltonian.

\subsection{Simulated density of states in the junction}
\label{sec:dos}

\begin{figure}[!h]
    \centering
        \includegraphics[width=0.6\textwidth]{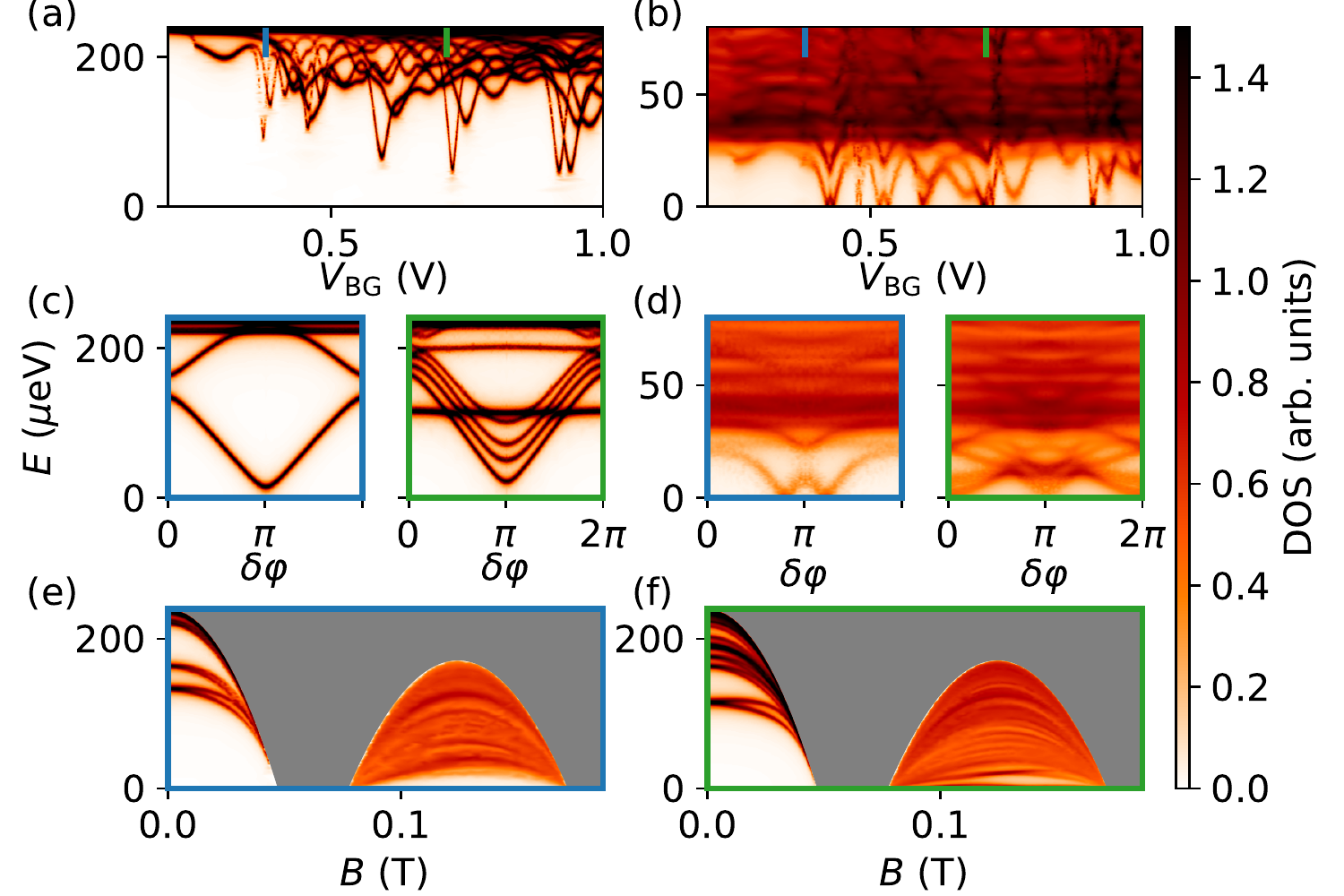}\vspace{-2mm}
    \caption[
    ]{\vspace{-0mm} (a,~b) DOS in the junction as a function of $V_\mathrm{BG}$ for $\delta\varphi=0$ at zero magnetic field ($n=0$) and at $B= 0.12$\,T ($n=1$). (c) DOS as a function of $\delta\varphi$ at zero magnetic field for $V_\mathrm{BG}=0.38$\,V and $V_\mathrm{BG}=0.71$\,V. (d) same as (c) but at $B=0.12\,$T in the middle of the first lobe. (g) DOS as a function of $V_\mathrm{BG}$ at $B=0.12\,$T for $\delta\varphi=0$. (e, ~f) $B$ dependence of DOS at $V_\mathrm{BG}=0.38$\,V and $V_\mathrm{BG}=0.71$\,V. The broadening used throughout this figure is $\eta=2$\,meV.
    }
    \label{fig:dos}
\end{figure}

The phase winding in the superconducting shell has dramatic effects on the ABS spectrum in the junction. In Figs.~\ref{fig:dos}(a,~b) we show the density of states (DOS) in the junction (obtained by integrating the LDOS from $x=-450$\,nm to $x=450$\,nm) in the zeroth lobe and in the first lobe as a function of $V_\mathrm{BG}$ at $\delta\varphi=0$. In the zeroth lobe, the energies of ABSs remain on the order of $\Delta$, whereas in the first lobe their energies can become much smaller. Note that in the first lobe, the bulk gap is on the order of 30\,$\mu$eV due to the presence of vortex states. If the junction is opened the gap is filled with low-energy ABSs. Only in the first lobe, states at energies much smaller than $\Delta$ at $\delta\varphi=0$ are observed. In Fig.~\ref{fig:dos}~(c,~d) we show the $\delta\varphi$ dependence of the DOS. Figure~\ref{fig:dos}(e,~f) presents the dependence on magnetic field of these ABSs at $\delta\varphi = 0$. They do not follow the same magnetic field dependence as the Al gap, often having their energy maximum at a different flux, which is also compatible with the experimental observations in Fig.~\ref{fig:Fireworks}(b) in the main text.

\subsection{Simulations in the trivial and topological regime}
\label{sec:top}

\begin{figure}
    \centering
        \includegraphics[width=0.6\textwidth]{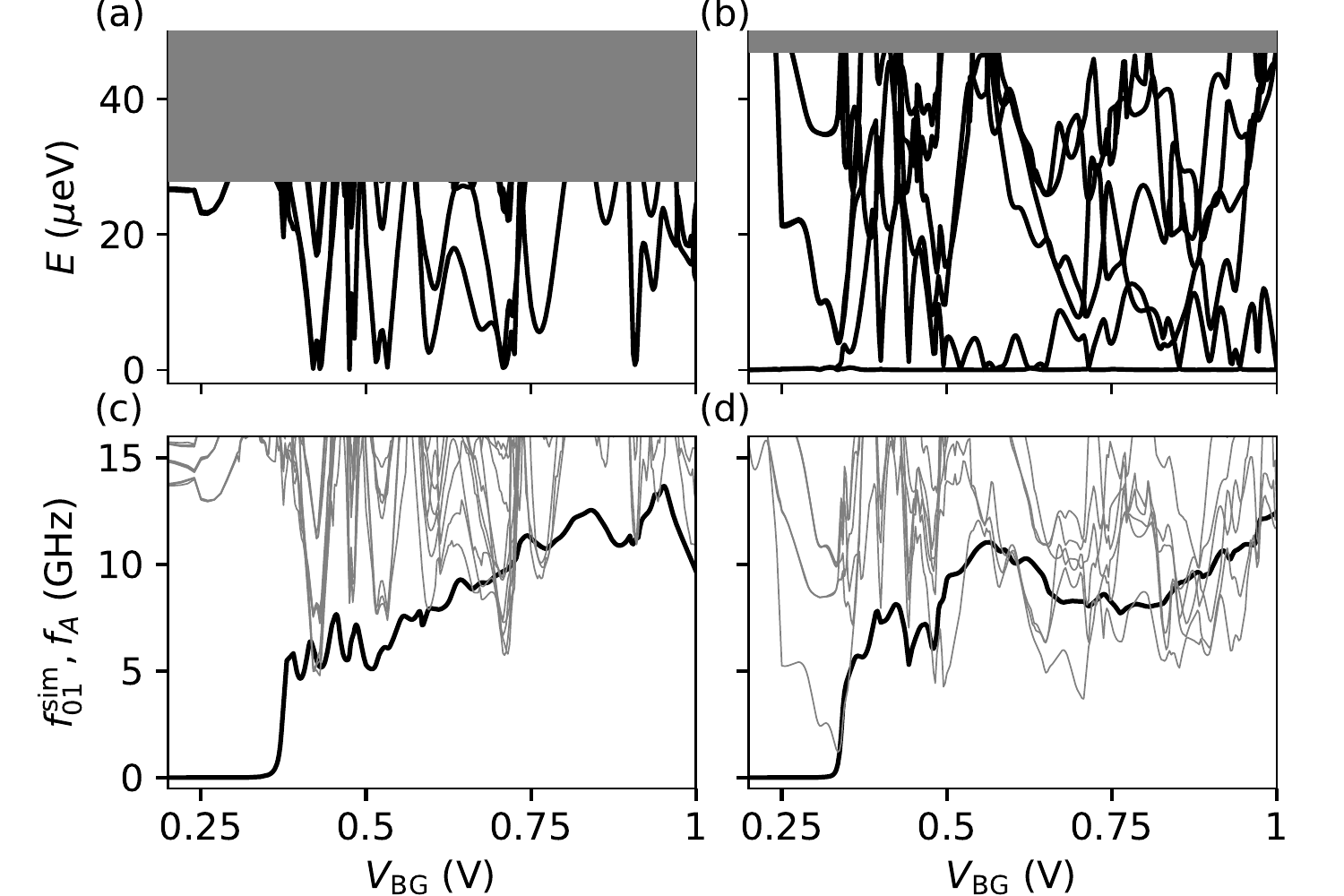}\vspace{-2mm}
    \caption[
    ]{\vspace{-0mm} (a,~b) Eigenenergies as a function of $V_\mathrm{BG}$ for $\delta\varphi=0$ in the first lobe ($B= 0.12$\,T, $n=1$) for parameters corresponding to the trivial phase ($U_0=140$\,meV, $\alpha=0$) and the topological phase ($U_0=150$\,meV, $\alpha=-0.1$\,eV\,nm). The continuum of bulk states is indicated in gray. (c,~d) Simulated qubit frequency $f_{01}^{\rm sim}$ (black) and two-quasiparticle transition frequencies $f_A$ (gray) as a function of $V_\mathrm{BG}$ for the same parameters as in (a,~b).
    }
    \label{fig:top}
\end{figure}

For certain parameters the full-shell NWs support a topological phase in the first lobe~\cite{Sole_Georg_2020}. In the main text, all results where generated in the trivial regime. In Fig.~\ref{fig:top}. we compare ABS energies and simulated qubit frequencies in the trivial and topological regime. In the topological case, transitions to the inert outer Majoranas were not considered for the two-quasiparticle transition frequencies $f_A$. Fig.~\ref{fig:top}(c) in the trivial case and Fig.~\ref{fig:top}(d) in the topological case yield qualitatively very similar spectra, making a distinction between the trivial and topological case difficult.

\end{document}